\documentclass{PoS}

\title{Heavy quarkonium production in hadronic collisions in TMD framework}

\author{Jian-Wei Qiu\\
Theory Center, Thomas Jefferson National Accelerator Facility, Newport News, VA 23606, USA\\
        E-mail: \email{jqiu@jlab.org}}

\author{\speaker{Kazuhiro Watanabe}
      \\
Physics Department, Old Dominion University, Norfolk, VA 23529, USA\\
Theory Center, Thomas Jefferson National Accelerator Facility, Newport News, VA 23606, USA\\
      E-mail: \email{watanabe@jlab.org}}

\abstract{
Heavy quarkonium ($\Upsilon$) production at low transverse momentum ($P_\perp$) in high-energy hadronic collisions is revisited from the point of view of transverse momentum dependent (TMD) framework. We perform resummation of double logarithmic correction associated with initial-state soft gluon shower for $b\bar b$ production by employing Collins-Soper-Sterman (CSS) formalism. We show that the CSS formalism provides a nice description of $\Upsilon$ production data in p+$\rm\bar{p}$ collisions at Tevatron and p+p collisions at the LHC.
}

\FullConference{QCD Evolution 2017\\
		 22-26 May, 2017\\
		 Jefferson Lab, Newport News, VA - USA}
\ShortTitle{Quarkonium production in hadronic collisions in TMD framework}

\begin{document}


\section{Introduction}

Transverse momentum ($P_\perp$) distribution of heavy quarkonium ($J/\psi$, $\Upsilon$) production in high-energy hadronic (p+p, p+$\rm{\bar{p}}$) collisions provides important playgrounds for studying perturbative QCD as well as nonperturbative hadronization dynamics at long distance. Quarkonium production at high $P_\perp\gg 2m\sim M_{H}$ with heavy quark mass $m$ and quarkonium mass $M_H$ can be evaluated in terms of QCD collinear factorization approach with perturbatively calculable hard coefficient functions along with universal fragmentation functions \cite{Kang:2014tta,Kang:2014pya}. When $P_\perp \sim 2m$, but, $\gg mv$ with heavy quark velocity $v$ in the pair's rest frame, non-relativistic QCD (NRQCD) factorization \cite{Bodwin:1994jh} could be a good approach to analyze heavy quarkonium production data \cite{Brambilla:2010cs}. 

On the other hand, neither collinear nor NRQCD factorization is expected to be valid in dealing with heavy quarkonium production at low $P_\perp\ll M_H$, since the measured $P_\perp$ could be sensitive to the nonperturbative intrinsic transverse momentum dependence of colliding partons, the nonperturbative scale $Q_s$ characterizing the multiple scattering with the incoming hadrons, and amount of transverse momentum of active partons generated by radiation (or shower) of gluons from the collisions.  Transverse momentum dependent (TMD) framework~\cite{Collins:2011zzd} has been a robust tool for studying observables with two scales, such as $P_\perp\ll M_H$.  In TMD framework, it is the factorized TMD parton distributions (TMDs) that take care of the various contributions to the active partons' transverse momentum.  If the effective value of the active partons' transverse momentum is dominated by {\it perturbatively calculable} contribution from the shower of soft gluons, and is much larger than the non-perturbative intrinsic parton transverse momentum as well as the scale generated by the multiple soft scattering, the TMD framework should have a good predictive power for the perturbatively calculated $P_\perp$ distribution of heavy quarkonium production.  

With the large heavy quark mass, especially for the bottom quark, we expect that the soft gluon radiation from the produced heavy quarks is much smaller than the radiation from the colliding light-partons, which could be strongly enhanced when $\sqrt{s}\gg M_H$.  Since the heavy quark pair production is dominated by the gluon-gluon fusion and gluons are much more likely to radiate than quarks, we conjecture that the $P_\perp$ distribution of $\Upsilon$ production in high-energy hadronic collisions is dominated by the leading double logarithmic contributions $\propto \alpha_s\ln^2(M_H^2/P_\perp^2)$ from the initial-state radiation, which can be summed to all-order in $\alpha_s$ using the Collins-Soper-Sterman (CSS) resummation formalism~\cite{Collins:1984kg}.  Early predictions of hadronic $\Upsilon$ production at Tevatron in the TMD framework were given in Ref.~\cite{Berger:2004cc,Berger:2004ct} (See also \cite{Sun:2012vc}), and were found to be consistent with data.
 Now the LHC experiments are running and providing rich data on quarkonium production. It is clearly meaningful and important to compare the theoretical calculations with the LHC data on the $P_\perp$ distribution of $\Upsilon$ production. This paper is aimed at presenting such comparisons.


\section{Transverse Momentum Dependent Framework}

In this paper, we shall employ the following expression for  differential cross section of inclusive $\Upsilon$ production in high-energy hadronic collisions ($A+B\rightarrow b\bar b+X\rightarrow \Upsilon+X$):
\begin{eqnarray}
\frac{d\sigma_{A+B\rightarrow {\Upsilon}+X}}{d^2P_\perp dy}=\int_{M_{\Upsilon}^2}^{(2M_B)^2} dM^2 F_{b\bar b\rightarrow \Upsilon}(M^2) \frac{d\sigma_{A+B\rightarrow {b\bar b}+X}}{dM^2 d^2P_\perp dy},
\label{eq:CEM}
\end{eqnarray}
where $M$ is invariant mass of a $b\bar b$ pair, $M_\Upsilon$ is the mass of $\Upsilon$, and $2M_B$ with $M_B=5.279\;$GeV is the $B\bar B$ decay threshold for the $\Upsilon$ system. $F_{b\bar b\rightarrow \Upsilon}$ is a transition probability to transform the $b\bar b$ pair to a $\Upsilon$ with a power-law distribution, and was fixed by the Tevatron data~\cite{Berger:2004cc}.  
The differential cross section for $b\bar b$ production in Eq.~(\ref{eq:CEM}) can be calculated within the framework of QCD collinear factorization, to obtain reliable predictions for the cross sections at high $P_\perp$.  A singular spike in the vicinity of $P_\perp=0_\perp$ indicates the failure of the fixed-order perturbative calculations for the low $P_\perp$ regime.
However, once we sum over large double logarithms $\alpha_s\ln^2(M^2/P_\perp^2)$ associated with initial-state gluon shower from incoming partons, the differential cross section for inclusive $b\bar b$ production remains finite at $P_\perp=0_\perp$, and can be written as 
\begin{eqnarray}
\frac{d\sigma_{A+B\rightarrow {b\bar b}+X}}{dM^2 d^2P_\perp dy}=\frac{d\sigma_{A+B\rightarrow{b\bar b}+X}}{dM^2d^2P_\perp dy}\bigg|_\textrm{\scriptsize resum}+\frac{d\sigma_{A+B\rightarrow{b\bar b}+X}}{dM^2d^2P_\perp dy}\bigg|_{\scriptsize Y}.
\end{eqnarray}
The first term in the right hand side of the above expression is the resummation term, and the second term is the so-called $Y$-term which is defined by
$\frac{d\sigma_{A+B\rightarrow{b\bar b}+X}}{dM^2d^2P_\perp dy}\big|_{\scriptsize Y}=\frac{d\sigma_{A+B\rightarrow{b\bar b}+X}}{dM^2d^2P_\perp dy}\big|_\textrm{\scriptsize pert}-\frac{d\sigma_{A+B\rightarrow{b\bar b}+X}}{dM^2d^2P_\perp dy}\big|_\textrm{\scriptsize asym}$.
The singularity arose in the perturbation term at $P_\perp=0_\perp$ and the other singularity in the resummation term around $P_\perp=M_\psi$ are canceled out by the asymptotic term. 
The perturbative and the asymptotic term could be obtained by fixed-order calculations in powers of $\alpha_s$, and  are available at next-to-leading order (NLO) at ${\cal O}(\alpha_s^3)$.

Following Collins-Soper-Sterman (CSS) resummation formalism in the impact parameter $b_\perp$ space~\cite{Collins:1984kg}, the resummation term can be written as
\begin{eqnarray}
\frac{d\sigma_{A+B\rightarrow b\bar b+X}}{dM^2dP_\perp^2 dy}=\int \frac{db_\perp}{2\pi}J_0(P_\perp b_\perp)\left(\sum_{q=u,d,s,c} b_\perp W_{q\bar q}\frac{d\hat{\sigma}_{q\bar q\rightarrow b\bar b}}{dM^2}+b_\perp W_{gg}\frac{d\hat{\sigma}_{gg\rightarrow b\bar b}}{dM^2}\right),
\label{eq:CSS-formalism}
\end{eqnarray}
where $J_0$ is a Bessel function,  and
$d\hat{\sigma}_{ij\rightarrow b\bar b}/dM^2$ are partonic cross sections for producing the $b\bar b$ pair with the leading-order (LO) contribution at ${\cal O}(\alpha_s^2)$ and high order corrections depending on the choice of factorization scheme.
The effective partonic flux in $b_\perp$-space, $W_{q\bar q}$ and $W_{gg}$ for $q\bar{q}$ and $gg$ channels, respectively, can be calculated perturbatively only at small $b_\perp$ below $b_{max}\sim 0.5\;$GeV$^{-1}$, 
\begin{eqnarray}
W^{perp}_{ij}(b_\perp,M,x_A,x_B)=e^{-S_{ij}(b_\perp,M)}f_{i/A}\left(x_A,\mu,\frac{c_0}{b_\perp}\right)f_{j/B}\left(x_B,\mu,\frac{c_0}{b_\perp}\right),
\label{eq:partonicflux}
\end{eqnarray}
where $x_{A,B}=M/\sqrt{s}e^{\pm y}$ are the longitudinal momentum fractions of incoming hadrons carried by colliding partons, $c_0=2e^{-\gamma_E}\sim1$ with Euler constant $\gamma_E$, and $S_{ij}$ (Sudakov factor), which sums up the leading soft and collinear contributions from the initial-state gluon shower, is given by
\begin{eqnarray}
S_{ij}(b_\perp,M)=\int_{c_0^2/b_\perp^2}^{M^2}\frac{d\bar \mu^2}{\bar\mu^2}\left[A_{ij}(\alpha_s)\ln\frac{M^2}{\bar\mu^2}+B_{ij}(\alpha_s)\right]
\end{eqnarray}
with perturbatively calculable coefficients $A_{ij}=\sum_{n=1}A_{ij}^{(n)}(\alpha_s/\pi)^n$ and $B_{ij}=\sum_{n=1}B_{ij}^{(n)}(\alpha_s/\pi)^n$. The modified parton distribution functions (PDFs) in Eq.~(\ref{eq:partonicflux}) are defined as
\begin{eqnarray}
f_{i/A}\left(x_A,\mu,\frac{c_0}{b_\perp}\right)=\sum_a\int_{x_A}^1\frac{d\xi}{\xi}\phi_{a/A}(\xi,\mu)C_{a\rightarrow i}\left(\frac{x_A}{\xi},\mu,\frac{c_0}{b_\perp}\right)
\end{eqnarray}
where $\phi_{a/A}(\xi,\mu)$ are the usual collinear PDFs with momentum fraction $\xi$ and factorization scale $\mu$. The coefficient $C_{a\rightarrow i}=\sum_{n=0}C_{a\rightarrow i}^{(n)}(\alpha_s/\pi)^n$ is also calculable perturbatively. Depending on the factorization scheme, the coefficients $A$, $B$, and $C$ could be process dependent, or made to be process independent by including the process-dependent contributions to the short-distance hard parts, $d\hat{\sigma}_{ij\rightarrow b\bar b}/dM^2$. In this work, we will keep only the process-independent terms for the coefficients of $A$, $B$, and $C$ in the power series in $\alpha_s$. Now it is worth emphasizing that for the $q\bar q$ channel the coefficients are the same as those of Drell-Yan production while the coefficients for the $gg$ channel are the same as those for Higgs boson production.

To derive the $P_\perp$ distribution in Eq.~(\ref{eq:CSS-formalism}), an extrapolation of the $b_\perp$-space distributions to the region where $b_\perp>b_{max}$ is required with the use of appropriate nonperturbative form factor as 
\begin{eqnarray}
W_{ij}(b_\perp,M,x_A,x_B)=W^{perp}_{ij}(b_{max},M,x_A,x_B)F_{ij}^{NP}(b_\perp,M,x_A,x_B;b_{max}).
\end{eqnarray}
In our calculations, we set $F_{ij}^{NP}=\exp\left[-\ln\left(\frac{M^2b_{max}^2}{c_0^2}\right)g_1(b_\perp^{2\alpha}-b_{max}^{2\alpha})\right]$ where two parameters $g_1$ and $\alpha$ are determined uniquely by requiring that $W_{ij}$ are continuous at $b_\perp=b_{max}$~\cite{Qiu:2000hf}, and see Ref.~\cite{Berger:2002ut} for other choices of $F_{ij}^{NP}$. We will see later that the nonperturbative form factor does not affect the $b\bar b$ production cross section because the perturbative distribution $W_{ij}^{perp}$ dominates in the $b_\perp$-space.


\section{Numerical results}

\begin{figure}
\centering
\includegraphics[width=0.45\textwidth]{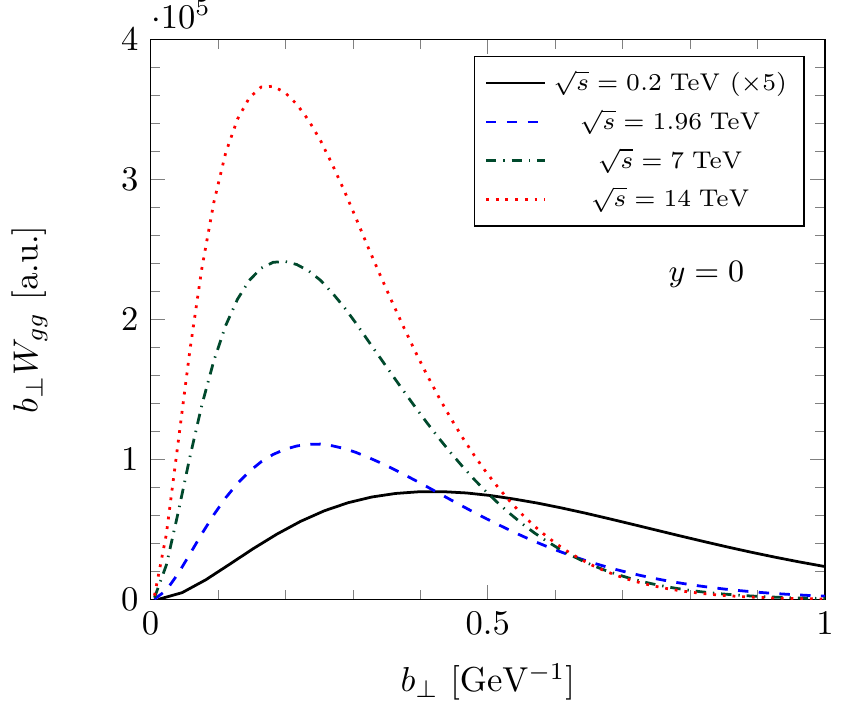}
\includegraphics[width=0.45\textwidth]{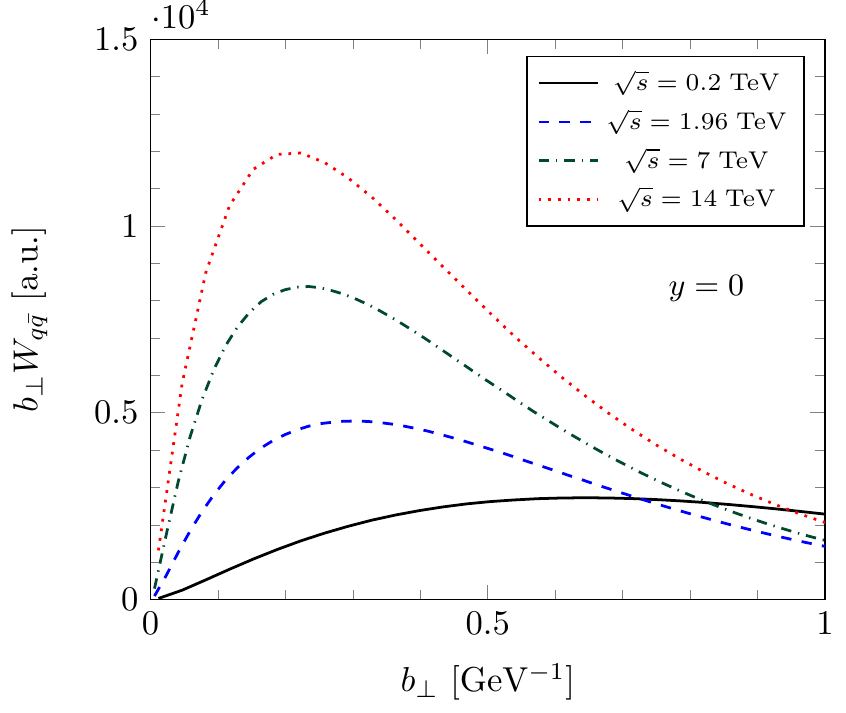}
\caption{The $b_\perp$ distribution of effective partonic flux at $y=0$, $W_{gg}$ (Left) and $W_{q\bar q}$ (Right), obtained by integrating over the range $2m_b<M<2M_B$ at several collider energies.}
\label{fig:wqq1800-y0}
\end{figure}

\begin{figure}
\centering
\includegraphics[width=0.6\textwidth]{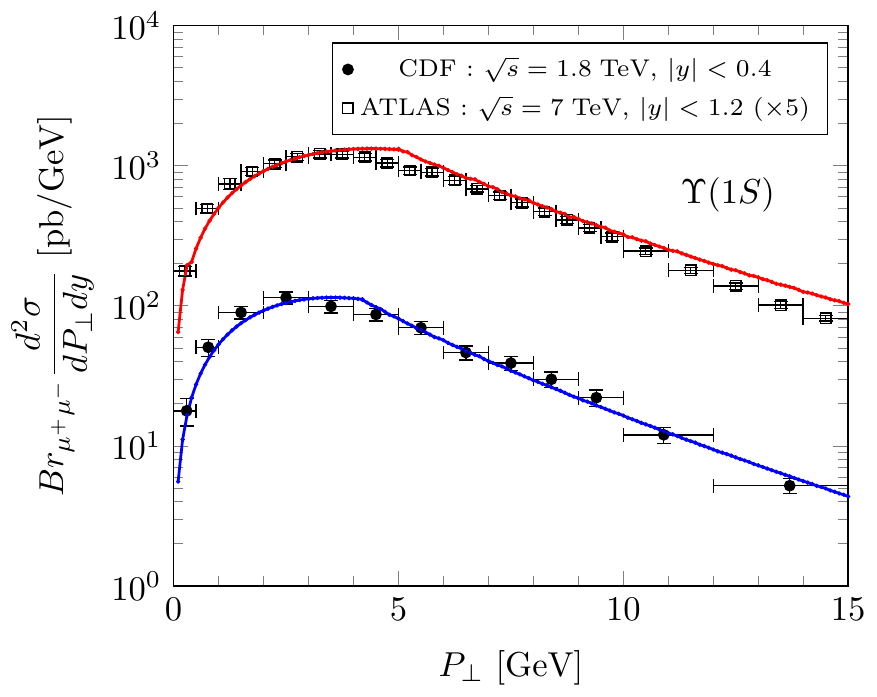}
\caption{Differential cross section for $\Upsilon(1S)$ production in hadronic collisions at Tevatron and the LHC in the middle rapidity region. All the input parameters are chosen to be the same as in Ref.~\cite{Berger:2004cc}. Data are taken from \cite{Acosta:2001gv,Aad:2012dlq}.}
\label{fig:xsection}
\end{figure}

We present in Figure~\ref{fig:wqq1800-y0} the $b_\perp$-distribution of $W_{gg}$ and $W_{q\bar q}$ in Eq.~(\ref{eq:CSS-formalism}) at various collider energies at the mid rapidity. We set $m_b=4.5\;$GeV and utilize CTEQ6M set for the collinear PDFs~\cite{Pumplin:2002vw}. We choose $\mu=c_0/b_\perp$ in the resummation term. For the $gg$ channel, one can find immediately that the peak position or the so-called saddle point~\cite{Qiu:2000hf} of $b_\perp W_{gg}$ shifts toward small-$b_\perp$ region as scattering energy $\sqrt{s}$ increases (more phase space for the gluon shower). At Tevatron and the LHC, the $b_\perp$-distribution of $W_{gg}$ is sharp and largely located at $b_\perp$ less than $b_{max}$. That is, we have a good predictive power for the $b\bar b$ production cross section calculated perturbatively without worrying about the uncertainty of the nonperturbative form factor at large $b_\perp$. On the other hand, the saddle point of $b_\perp W_{gg}$ is located around $b_\perp\sim b_{max}$ at RHIC energy ($\sqrt{s}=0.2\;$TeV). Therefore the nonperturbative form factor could play an essential role for calculating the $b\bar b$ cross section at RHIC energy. For the $q\bar q$ channel, the $b_\perp$-distribution of $W_{q\bar q}$ is more broad so that the nonperturbative form factor is more relevant. Nevertheless, in our calculations, we do not need to worry too much about it because the size of contribution from the $gg$ channel is more than an order of magnitude larger than that from the $q\bar{q}$ channel.

Figure~\ref{fig:xsection} displays differential cross sections for $\Upsilon(1S)$ production in hadronic collisions at Tevatron and the LHC by computing Eq.~(\ref{eq:CSS-formalism}) with Eq.~(\ref{eq:CEM}). We set $\mu=0.5\sqrt{M^2+P_\perp^2}$ for the perturbation term. At Tevatron, we reproduce the early prediction in Ref.~\cite{Berger:2004cc} by setting $F_{b\bar b\rightarrow\Upsilon}=C_\Upsilon=0.044$ that was obtained by data fitting in Ref.~\cite{Berger:2004cc}, which is effectively a Color-Evaporation-Model calculation~\cite{Brambilla:2010cs}. To compare with data, we simply switch the resummation term to the NLO perturbative term at the intersection of two curves around $P_\perp\sim M_\Upsilon/2$, instead of using the $Y$-term. We have also multiplied the resummation term by a factor $K_{r}=1.22$ to match the perturbation result at the intersection. At the LHC, there is more phase space for gluons shower, and we expect our predictions with the same parameters set to be consistent with the data, which is confirmed nicely by the data up to around $P_\perp=10\;$GeV. It is worth noting that the matching point shifts toward larger $P_\perp$ at the LHC compared to that at Tevatron. This is because an increase in the scattering energy allows more phase space for  incoming partons to radiate.


\section{Summary}

We have performed numerical calculations for $\Upsilon$ production in high-energy hadronic collisions in terms of the Collins-Soper-Sterman resummation formalism in the TMD framework. The behavior of $W_{gg}$ and $W_{q\bar q}$ in the $b_\perp$-space at Tevatron and the LHC clearly shows that our perturbatively calculated results are reliable without much ambiguities associated with the nonperturbative Sudakov factor at large $b_\perp$. Our results can naturally describe both the Tevatron data and the LHC data for $P_\perp$ spectrum of hadronic $\Upsilon(1S)$ production at low $P_\perp$ by keeping all the parameters the same.  Fixing the parameters $g_1$ and $\alpha$ by the continuity of the $b_\perp$-space distribution takes care of the potential $\sqrt{s}$-dependence of the nonperturbative form factor  $F_{ij}^{NP}$ dynamically.

It should be an interesting project to address $\Upsilon$ production in proton-nucleus (p+A) collisions especially at forward rapidity where the multiple scattering effect with a larger $Q_s$ for a heavy nucleus could be as important as the initial-state gluon shower, as pointed out in Ref.~\cite{Qiu:2013qka}. The Sudakov effect from gluon shower on top of the multiple scattering effect for $\Upsilon$ production in p+p and p+A collisions was examined in small-$x$ color-glass-condensate framework~\cite{Watanabe:2015yca}.  It was found~\cite{Watanabe:2015yca} that the Sudakov effect is predominant over the multiple scattering effect, which is encoded in the saturation scale, in p+p collisions, while these two effects could be comparable in p+A collisions. We leave the extension of our work to p+A collisions for a future study.

\subsection*{Acknowledgement}
We thank Z.-B.~Kang, Y.-Q.~Ma, B.-W.~Xiao, and F.~Yuan for useful discussions and valuable collaborations on related topics. This work is supported by Jefferson Science Associates, LLC under  U.S. DOE Contract No.~DE-AC05-06OR23177 and by U.S. DOE Grant No.~DE-FG02-97ER41028.

\end{document}